\documentclass{article}
\usepackage[T1]{fontenc}
\usepackage[utf8]{inputenc}

\usepackage{dcase2024,amsmath,graphicx,times,booktabs, tabularx,bm,amssymb}
\usepackage{comment}
\usepackage[hyphens]{url}
\usepackage{hyperref}

\title{Description and Discussion on DCASE 2025 Challenge Task 4:\\ Spatial Semantic Segmentation of Sound Scenes}

\makeatletter
\def\thirdlinename#1{\gdef\@thirdlinename{{\em #1}\\}}
\gdef\@thirdlinename{}

\renewcommand{\@maketitle}{\newpage
  \null
  \vskip 1em
  \begin{center}
    {\large\bf \@title \par}
    \vskip 1.5em
    {\large\lineskip .5em
     \begin{tabular}[t]{c}
       \@name
       \ifx\@secondlinename\@empty\else \\[-1em]\@secondlinename\fi
       \ifx\@thirdlinename \@empty\else \\[-1em]\@thirdlinename \fi
       \\ \@address
     \end{tabular}\par}
  \end{center}
  \vskip 1.5em}
\makeatother

\name{Masahiro Yasuda$^{1}$,
      Binh Thien Nguyen$^{1}$, 
      Noboru Harada$^{1}$,
      Romain Serizel$^{2}$,
      Mayank Mishra$^{2}$}

\secondlinename{Marc Delcroix$^{1}$,
                Shoko Araki$^{1}$,
                Daiki Takeuchi$^{1}$,
                Daisuke Niizumi$^{1}$}

\thirdlinename{Yasunori Ohishi$^{1}$,
               Tomohiro Nakatani$^{1}$,
               Takao Kawamura$^{3}$,
               Nobutaka Ono$^{3}$}

\address{$^1$ NTT Corporation, Japan, masahiro.yasuda@ntt.com\\          
        $^2$  Universit\'e de Lorraine, France\\
        $^3$  Tokyo Metropolitan University, Japan\\}

\begin{document}
\ninept
\maketitle

\begin{abstract}
Spatial Semantic Segmentation of Sound Scenes (S5) aims to enhance technologies for sound event detection and separation from multi-channel input signals that mix multiple sound events with spatial information. This is a fundamental basis of immersive communication.
The ultimate goal is to separate sound event signals with 6 Degrees of Freedom (6DoF) information into dry sound object signals and metadata about the object type (sound event class) and representing spatial information, including direction. However, because several existing challenge tasks already provide some of the subset functions, this task for this year focuses on detecting and separating sound events from multi-channel spatial input signals.
This paper outlines the S5 task setting of the Detection and Classification of Acoustic Scenes and Events (DCASE) 2025 Challenge Task 4 and the DCASE2025 Task 4 Dataset, newly recorded and curated for this task. 
We also report experimental results for an S5 system trained and evaluated on this dataset.
The full version of this paper will be published after the challenge results are made public.
\end{abstract}


\section{Introduction}
\label{sec:intro}
This paper summarizes a newly introduced task to the Detection and Classification of Acoustic Scenes and Events (DCASE) 2025 challenge, named Task 4: Spatial Semantic Segmentation of Sound Scenes (S5)~\cite{dcase2025t4web}, and discusses the results of the challenge. 

This task requires systems to detect and extract sound events from multi-channel spatial input signals.
The input signal contains, at most, three simultaneous sound events plus optional non-directional
background noise. Each output signal should contain one isolated sound event with a predicted label for the event class.

\begin{figure*}[t]
  \centering
  \centerline{\includegraphics[width=2\columnwidth]{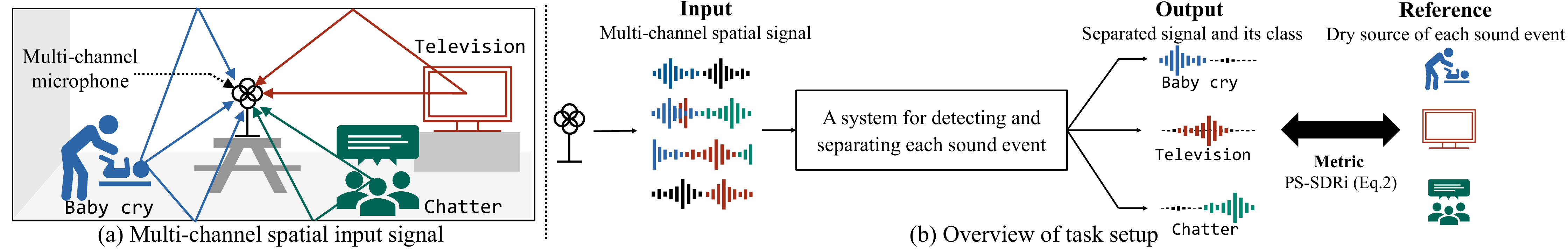}}
  \caption{Overview of spatial semantic segmentation of sound scenes.}
 \label{fig:tasksetup}
\end{figure*}

To enable immersive communication (including XR), a complex sound scene should be separated into sound objects representing sound events. 
We want to investigate how we could utilize directional information and prior knowledge of the semantic information of the event class to enable spatial semantic segmentation of sound scenes. 
As the first step for this year, the systems must detect contained sound events and separate those sound events into dry
sources from multi-channel spatial input signals.

Unlike DCASE 2021 Task 4: Sound Event Detection and Separation in Domestic
Environments~\cite{dcase2021t4, dcase2021t4web}, the separation/extraction should be performed from a multi-channel signal that contains spatial information. DCASE 2024 Task 3: Audio and Audiovisual Sound Event Localization and Detection with Source Distance Estimation~\cite{dcase2024t3web} provides
complementary functions generating meta-information for Direction of Arrival (DoA). Thus, this task carefully avoids the DoA part but requires spatial sound separation, which is a mandatory aspect of this task.
\section{Task setting of S5 [5]}
\label{sec:tasksetting}
The S5 task, originally proposed in our prior work~\cite{eusipco_s5}, aims to detect and separate the sounds of each sound event from signals observed by a multi-channel microphone at various locations in a real environment. This section introduces the notation and task settings for the S5 task.

Let $\bm{Y} = [\bm{y}^{(1)},\dots, \bm{y}^{(M)}]^\top \in \mathbb{R}^{M \times T}$ be the multi-channel time-domain mixture signal of length $T$, recorded with an array of $M$ microphones, where $\{\cdot\}^\top$ is the matrix transposition.
We denote $C=\{c_1, ...,c_K\}$ the set of source labels in the mixture, where the source count $K$ can vary from $1$ to $K_\textrm{max}$.
The $m$-th channel of $\bm{Y}$ can be modeled as
\begin{equation}\label{eq:mixture}
	\bm{y}^{(m)}=\sum_{k=1}^{K} \bm{h}^{(m)}_k*\bm{s}_k + \bm{n}^{(m)} 
	            =\sum_{k=1}^{K}\bm{x}^{(m)}_k + \bm{n}^{(m)},
\end{equation}
where $\bm{s}_k\in\mathbb{R}^T$ is the single-channel dry source signal corresponding to the label $c_k$, $\bm{h}^{(m)}_k \in \mathbb{R}^H$ is the $m$-th channel of the length-$H$ room impulse response (RIR) at the spatial position of $\bm{s}_k$, and $\bm{n}^{(m)}\in\mathbb{R}^T$ is the $m$-th channel of the multi-channel noise signal.

In the S5 task, the desired outputs are the individual sound events $\{\bm{s}_1,\ldots,\bm{s}_K\}$ themselves. 
However, directly recovering $\bm{s}_k$ from the mixtures $\bm{y}^{(m)}$ would require compensating for propagation delays that vary with the source–microphone distance. This additional challenge lies outside S5's core focus on sound event detection and separation. 
Hence, as a reasonable relaxation of the task, we instead define the targets as the source signals convolved with the direct-path impulse response toward a designated reference microphone.

\section{DCASE2025 Task4 Dataset}
\subsection{General overview}
For the S5 task, we designed and recorded a new dataset named \textbf{DCASE2025 Task4 Dataset}~\cite{dcase2025t4devset,dcase2025t4evalset}.
Given the task definition in Sec.~\ref{sec:tasksetting}, a dataset suitable for S5 requires the following resources:
\begin{itemize}
  \item \textbf{Isolated target sound events $\bm{s}_k$:} one-shot recordings of diverse sound event classes.  In light of Eq.~\eqref{eq:mixture}, these signals are preferably captured in anechoic conditions.
  \item \textbf{Room-impulse responses (RIRs) $\bm{h}^{(m)}_k$:} multichannel RIRs measured in various rooms.
  \item \textbf{Environmental noise $\bm{n}^{(m)}$:} environmental noise recorded with the same multichannel microphone array used for the RIRs.
\end{itemize}
In addition to stationary, direction-independent environmental noise, it is often useful to include sporadic sound events that do not belong to the target classes.  
We refer to these as \textit{interference sounds} and handle them as follows:
\begin{itemize}
  \item \textbf{Interference sounds:} Recording of sound events of classes not included in the isolated target sound events mentioned above.
  When forming mixtures, the interference signals are processed similarly to the target sound events.
\end{itemize}

In order to meet these requirements for S5 tasks, the DCASE2025 Task4 Dataset consists of new recordings and curated data from publicly available datasets~\cite{fsd50k, semhear, foameir}.
Eighteen (18) classes are selected for the target sound events: ``AlarmClock", ``BicycleBell", ``Blender", ``Buzzer", ``Clapping", ``Cough", ``CupboardOpenClose", ``Dishes", ``Doorbell", ``FootSteps", ``HairDryer", ``MechanicalFans", ``MusicalKeyboard", ``Percussion", ``Pour", ``Speech", ``Typing", and ``VacuumCleaner".
Among the materials that make up the DCASE2025 Task4 Dataset, our newly recorded material and curated materials from FOA-MEIR~\cite{foameir} are released on Zenodo~\cite{dcase2025t4devset, dcase2025t4evalset}.
The remaining materials are obtained and filtered from their respective public datasets via a download script that we provide on GitHub~\cite{dcase2025t4github}.

The training and evaluation mixtures $\bm{y}^{(m)}$ were synthesized from the above materials using a modified version of a spatial-audio simulator named SpatialScaper~\cite{spatialscaper}. 
All mixtures were synthesized at 32kHz/16bit. The source materials and rendering parameters are described in the following section.

\subsection{The development dataset}
The development dataset of the DCASE2025 Task4 Dataset~\cite{dcase2025t4devset} was constructed from various datasets, including both existing and newly recorded data specifically for this task.
The isolated target sound events for the development set consist of our newly recorded data and curated samples from FSD50K~\cite{fsd50k} and EARS~\cite{ears}.
RIR dataset was constructed by merging newly recorded RIRs for this task with curated material from the publicly released FOA-MEIR dataset~\cite{foameir}.
All environmental-noise signals were curated from the publicly available FOA-MEIR dataset~\cite{foameir}. All interference sounds were curated from the publicly available dataset. This curation was performed by removing items related to our target 18 classes from the background set in the dataset for Semantic Hearing~\cite{semhear}.
The procedures for the new recordings and the curated material drawn from public datasets are detailed in Sec.~\ref{sec:recording}.

The development dataset was further divided into three subsets: training, validation, and test. 
Mixtures for the test subset were pre-synthesized, while for the training and validation subsets, participants may generate mixtures using the provided data.
All mixture clips are fixed at 10 seconds. 
Each clip contains between one and three sound events, with at most three events active simultaneously. 
The RIR assigned to each sound event is selected randomly, subject to the constraint that all RIRs within a mixture must be from the same microphone position in the same room.
The SNR for each target sound event ranges from 5 to 20 dB, with respect to the background environmental noise.
Each mixture also includes one or two interfering sound events, with SNRs ranging from 0 to 15 dB.
The RIR assigned to each sound event is selected from the RIR dataset with the constraint that, within a mixture, all RIRs must originate from the same microphone position.

\subsection{The evaluation dataset}
To ensure fairness in the challenge, isolated sound events, RIR, interference sounds, and environmental noise were all newly recorded for the evaluation data. In other words, the evaluation data does not include any publicly available data.

A later version of this report will provide further details of the DCASE2025 Task4 Dataset, along with the experimental analysis results.
The evaluation dataset is available at Zenodo~\cite{dcase2025t4evalset}.

\subsection{Recording details}
\label{sec:recording}
\subsubsection{Recording details of isolated sound events}

\begin{figure}[t]
  \centering
\centerline{\includegraphics[width=\columnwidth]{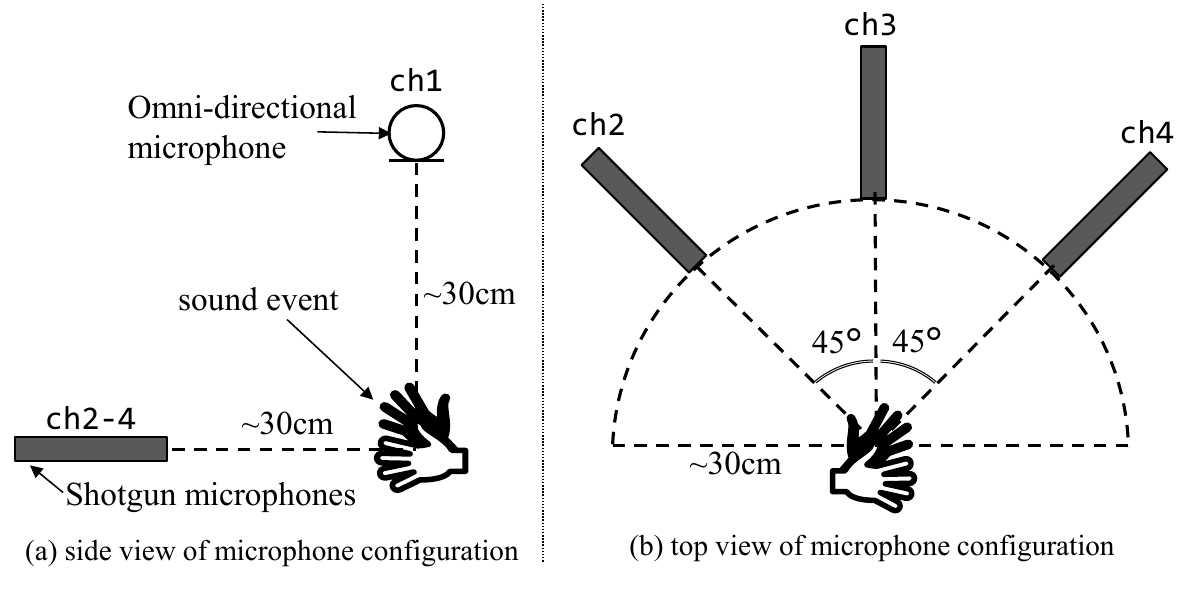}}
  \caption{The Configuration of microphones used to record isolated sound events in an anechoic chamber}
 \label{fig:micsetup}
\end{figure}

Isolated target sound events were recorded in an anechoic chamber using microphones. Fig.~\ref{fig:micsetup} shows the configuration of microphones for this recording. The recording was made using three cardioid microphones to capture the sound events from the left, front, and right, and one omnidirectional microphone to capture the sound from above. In the S5 task, it is assumed that you will simply select a single channel (e.g., ch=3) from these and use it as a monaural sound event. 

\subsubsection{Recording details of room impulse response}

All RIRs were captured with the same first-order Ambisonics microphone (Sennheiser AMBEO VR Mic) and are provided in B-format (AmbiX). The new recordings comprise 540 RIRs measured at five positions distributed across three acoustically distinct rooms. At each position, 108 RIRs were gathered while systematically varying the source-microphone geometry:
(i) the azimuth was swept in $20^{\circ}$ steps to cover the full $360^{\circ}$;
(ii) the elevation was set to $-20^{\circ}$, $0^{\circ}$, or $20^{\circ}$; and
(iii) the source-microphone distance was chosen from the range $0.75$--$1.50$~m.

\section{Evaluation method, metric [5]}

As evaluation metrics for the S5 task, we used class-aware signal-to-distortion ratio improvement (CA-SDRi) proposed for the S5 task~\cite{eusipco_s5}.
The key concept of CA-SDRi is that the estimated and reference sources are aligned by their labels, with the waveform metric being calculated when the label is correctly predicted, i.e., $c_k \in C \cap \hat{C}$.
In cases of incorrect label prediction, penalty values are accumulated.
As metrics with these properties, CA-SDRi metric is defined as
\begin{equation}
\begin{split}
	\textrm{CA-SDRi}&\left(\{\hat{\bm{x}}_1, \dots,\hat{\bm{x}}_{\hat{K}}\}, \{\bm{x}_1, \dots,\bm{x}_K\}, \hat{C}, C, \bm{y}\right) \\ &= \frac{1}{| C \cup \hat{C} |} \sum_{c_k \in C \cup \hat{C}} P_{c_k},
\end{split}
\end{equation}
where $| C \cup \hat{C} |$ is the length of the set union. The metric component $P_{c_k}$ is calculated as
\begin{equation}
P_{c_k\in C \cup \hat{C}} = 
\begin{cases}
	\textrm{SDRi}(\hat{\bm{x}}_k, \bm{x}_k, \bm{y}), &\text{if } c_k \in C \cap \hat{C}\\
	\mathcal{P}^\textrm{FN}_{c_k}, &\text{if } c_k \in C \text{ and } c_k \notin \hat{C}\\
	\mathcal{P}^\textrm{FP}_{c_k}, &\text{if } c_k \notin C \text{ and } c_k \in \hat{C}
\end{cases},
\end{equation}
where the SDRi is calculated as
\begin{equation}
	\textrm{SDRi}(\hat{\bm{x}}_k, \bm{x}_k, \bm{y})
	= \textrm{SDR}(\hat{\bm{x}}_k, \bm{x}_k) - \textrm{SDR}(\bm{y} , \bm{x}_k),
\end{equation}
\begin{equation}\label{sq:sdr}
	\textrm{SDR}(\hat{\bm{x}}, \bm{x}) 
	= 10\log_{10} \left( \frac{\|\bm{x}\|^2}{\|\bm{x} - \hat{\bm{x}}\|^2} \right).
\end{equation}
$\mathcal{P}^\textrm{FN}_{c_k}$ and $\mathcal{P}^\textrm{FP}_{c_k}$ are the penalty values for false negative (FN) and false positive (FP), both set to $0$, indicating that incorrect predictions do not contribute to any improvement in the metric.

Finally, the ranking scores are derived from the averages for each clip of CA-SDRi. In addition to the primary ranking score function CA-SDRi, we will also provide PESQ~\cite{pesq} and STOI~\cite{stoi} for speech, and PEAQ scores for signals other than speech as informative metrics representing perceptual quality.

\section{Experimental results and discussion}

This section presents the performance of the baseline systems, ResUNet and ResUNetK, proposed 
in~\cite{eusipco_s5}.
Both are two-stage systems that share the same first stage: a masked modeling duo (M2D)\cite{m2d}-based audio tagging (AT) model used for sound event classification. In the second stage, sound source separation is performed using two different approaches. ResUNet, adopted the model introduced in\cite{uss}, extracts a single source for each detected event. In contrast, ResUNetK extends this by allowing concurrent querying of multiple sources. The code for the baseline is available on GitHub~\cite{dcase2025t4baseline}.

Table 1 shows classification accuracies and CA-SDRi scores of the baseline systems on the Evaluation dataset and the test subset of the Development dataset.
These results are consistent with those reported in \cite{eusipco_s5}, where ResUNetK outperformed ResUNet on both datasets, despite using the same classification model.
In addition, the CA-SDRi is shown to vary proportionally with classification accuracy, further validating its efficacy in simultaneously evaluating both source separation and class label prediction in the S5 task.
\begin{table}
    \centering
\resizebox{\linewidth}{!}{
    \begin{tabular}{lccccc} \hline
    System & \multicolumn{2}{c}{Evaluation set} & \multicolumn{2}{c}{Development set} \\  
           & Ranking Score (CA-SDRi) $\uparrow$ & Acc. $\uparrow$ & CA-SDRi $\uparrow$ & Acc.  $\uparrow$ \\\hline
    Nguyen\_NTT\_task4\_1 (ResUNetK) & $6.60$ & $51.48$ & $11.09$ & $59.80$ \\
    Nguyen\_NTT\_task4\_2 (ResUNet) & $5.72$ & $51.48$ & $11.03$ & $59.80$  \\\hline 
    \end{tabular}}
    \caption{Experimental results of baselines.}
    \label{tab:my_label2}
\end{table}

\section{ACKNOWLEDGMENT}
\label{sec:ack}
This work was partially supported by JST Strategic International Collaborative Research Program (SICORP), Grant Number JPMJSC2306, Japan.

This work was partially supported by the Agence Nationale de la Recherche (Project Confluence, grant number ANR-23-EDIA-0003).

\bibliographystyle{IEEEtran}
\bibliography{refs}

\begin{thebibliography}{10}
\providecommand{\url}[1]{#1}
\def\UrlFont{\rmfamily}
\providecommand{\newblock}{\relax}
\providecommand{\bibinfo}[2]{#2}
\providecommand\BIBentrySTDinterwordspacing{\spaceskip=0pt\relax}
\providecommand\BIBentryALTinterwordstretchfactor{4}
\providecommand\BIBentryALTinterwordspacing{\spaceskip=\fontdimen2\font plus
\BIBentryALTinterwordstretchfactor\fontdimen3\font minus \fontdimen4\font\relax}
\providecommand\BIBforeignlanguage[2]{{%
\expandafter\ifx\csname l@#1\endcsname\relax
\typeout{** WARNING: IEEEtran.bst: No hyphenation pattern has been}%
\typeout{** loaded for the language `#1'. Using the pattern for}%
\typeout{** the default language instead.}%
\else
\language=\csname l@#1\endcsname
\fi
#2}}

\bibitem{dcase2025t4web}
\BIBentryALTinterwordspacing
(2025) {DCASE2025} challenge task 4: Spatial semantic segmentation of sound scenes. [Online]. Available: \url{https://dcase.community/challenge2025/task-spatial-semantic-segmentation-of-sound-scenes}
\BIBentrySTDinterwordspacing

\bibitem{dcase2021t4}
\BIBentryALTinterwordspacing
N.~Turpault, R.~Serizel, A.~Parag~Shah, and J.~Salamon, ``{Sound event detection in domestic environments with weakly labeled data and soundscape synthesis},'' in \emph{{Workshop on Detection and Classification of Acoustic Scenes and Events}}, New York City, United States, October 2019. [Online]. Available: \url{https://hal.inria.fr/hal-02160855}
\BIBentrySTDinterwordspacing

\bibitem{dcase2021t4web}
\BIBentryALTinterwordspacing
(2021) {DCASE2021} challenge task 4: Sound event detection and separation in domestic environments. [Online]. Available: \url{https://dcase.community/challenge2021/task-sound-event-detection-and-separation-in-domestic-environments}
\BIBentrySTDinterwordspacing

\bibitem{dcase2024t3web}
\BIBentryALTinterwordspacing
(2024) {DCASE2024} challenge task 3: Audio and audiovisual sound event localization and detection with source distance estimation. [Online]. Available: \url{https://dcase.community/challenge2024/task-audio-and-audiovisual-sound-event-localization-and-detection-with-source-distance-estimation}
\BIBentrySTDinterwordspacing

\bibitem{eusipco_s5}
\BIBentryALTinterwordspacing
B.~T. Nguyen, M.~Yasuda, D.~Takeuchi, D.~Niizumi, Y.~Ohishi, and N.~Harada, ``Baseline systems and evaluation metrics for spatial semantic segmentation of sound scenes,'' in \emph{2025 33rd European Signal Processing Conference (EUSIPCO)}, 2025. [Online]. Available: \url{https://arxiv.org/abs/2503.22088}
\BIBentrySTDinterwordspacing

\bibitem{dcase2025t4devset}
\BIBentryALTinterwordspacing
(2025) {DCASE}2025 {T}ask 4 {S5} {D}evelopment set. [Online]. Available: \url{https://zenodo.org/records/15117227}
\BIBentrySTDinterwordspacing

\bibitem{dcase2025t4evalset}
\BIBentryALTinterwordspacing
(2025) {DCASE}2025 {T}ask 4 {S5} {E}valuation set. [Online]. Available: \url{https://zenodo.org/records/15553984}
\BIBentrySTDinterwordspacing

\bibitem{fsd50k}
E.~Fonseca, X.~Favory, J.~Pons, F.~Font, and X.~Serra, ``{FSD50K}: an open dataset of human-labeled sound events,'' \emph{IEEE/ACM Transactions on Audio, Speech, and Language Processing}, vol.~30, pp. 829--852, 2021.

\bibitem{semhear}
B.~Veluri, M.~Itani, J.~Chan, T.~Yoshioka, and S.~Gollakota, ``Semantic hearing: Programming acoustic scenes with binaural hearables,'' in \emph{Proceedings of the 36th Annual ACM Symposium on User Interface Software and Technology}, 2023, pp. 1--15.

\bibitem{foameir}
M.~Yasuda, Y.~Ohishi, and S.~Saito, ``Echo-aware adaptation of sound event localization and detection in unknown environments,'' in \emph{IEEE Intl. Conf. on Acoust., Speech \& Sig. Proc. (ICASSP)}, 2022, pp. 226--230.

\bibitem{dcase2025t4github}
\BIBentryALTinterwordspacing
(2025) {DCASE2025 Task 4 Baseline}. [Online]. Available: \url{https://github.com/nttcslab/dcase2025_task4_baseline}
\BIBentrySTDinterwordspacing

\bibitem{spatialscaper}
I.~R. Roman, C.~Ick, S.~Ding, A.~S. Roman, B.~McFee, and J.~P. Bello, ``Spatial scaper: a library to simulate and augment soundscapes for sound event localization and detection in realistic rooms,'' in \emph{IEEE Intl. Conf. on Acoust., Speech \& Sig. Proc. (ICASSP)}, 2024, pp. 1221--1225.

\bibitem{ears}
J.~Richter, Y.-C. Wu, S.~Krenn, S.~Welker, B.~Lay, S.~Watanabe, A.~Richard, and T.~Gerkmann, ``Ears: An anechoic fullband speech dataset benchmarked for speech enhancement and dereverberation,'' in \emph{Interspeech 2024}, 2024, pp. 4873--4877.

\bibitem{pesq}
A.~Rix, J.~Beerends, M.~Hollier, and A.~Hekstra, ``Perceptual evaluation of speech quality (pesq)-a new method for speech quality assessment of telephone networks and codecs,'' in \emph{2001 IEEE International Conference on Acoustics, Speech, and Signal Processing. Proceedings (Cat. No.01CH37221)}, vol.~2, 2001, pp. 749--752 vol.2.

\bibitem{stoi}
C.~H. Taal, R.~C. Hendriks, R.~Heusdens, and J.~Jensen, ``A short-time objective intelligibility measure for time-frequency weighted noisy speech,'' in \emph{2010 IEEE International Conference on Acoustics, Speech and Signal Processing}, 2010, pp. 4214--4217.

\bibitem{m2d}
D.~Niizumi, D.~Takeuchi, Y.~Ohishi, N.~Harada, and K.~Kashino, ``Masked modeling duo: Towards a universal audio pre-training framework,'' \emph{IEEE/ACM Trans. on Audio, Speech, and Lang. Process.}, 2024.

\bibitem{uss}
Q.~Kong, K.~Chen, H.~Liu, X.~Du, T.~Berg-Kirkpatrick, S.~Dubnov, and M.~D. Plumbley, ``Universal source separation with weakly labelled data,'' \emph{arXiv preprint arXiv:2305.07447}, 2023.

\bibitem{dcase2025t4baseline}
\BIBentryALTinterwordspacing
(2025) Github: nttcslab/dcase2025\_task4\_baseline. [Online]. Available: \url{https://github.com/nttcslab/dcase2025_task4_baseline}
\BIBentrySTDinterwordspacing

\end{thebibliography}

\end{document}